\documentclass[preprint2]{emulateapj}

\usepackage{graphicx}
\usepackage{natbib}
\usepackage{amsmath}
\usepackage[version=3]{mhchem} 

\shorttitle{Abiotic O2 atmospheres on terrestrial planets}
\shortauthors{Wordsworth et al.}

\begin{document}


\title{Abiotic oxygen-dominated atmospheres on terrestrial habitable zone planets}


\author{Robin Wordsworth and Raymond Pierrehumbert\\
\vspace{0.1in}
\normalsize{5734 South Ellis Avenue, Department of the Geophysical Sciences}\\
\normalsize{University of Chicago, Chicago, IL 60622, USA}\\
}

\begin{abstract}
Detection of life on other planets requires identification of biosignatures, i.e., observable planetary properties that robustly indicate the presence of a biosphere. One of the most widely accepted biosignatures for an Earth-like planet is an atmosphere where oxygen is a major constituent. Here we show that lifeless habitable zone terrestrial planets around any star type may develop oxygen-dominated atmospheres as a result of water photolysis, because the cold trap mechanism that protects \ce{H2O} on Earth is ineffective when the atmospheric inventory of non-condensing gases (e.g., \ce{N2}, \ce{Ar}) is low. Hence the spectral features of \ce{O2} and \ce{O3} alone cannot be regarded as robust signs of extraterrestrial life.
\end{abstract}



\maketitle

\section{Introduction}

The rapid growth of exoplanet discovery and characterization over the last two decades has fueled hopes that in the relatively near future, we may be able to observe the atmospheres of Earth-like planets spectroscopically. Such targets will be intrinsically interesting for comparative planetology, but also for the major reason that they may host life. To search for life on exoplanets by observing their atmospheres, we must first decide on spectral features that can be used as biosignatures. Despite extensive theoretical study of various possibilities, detections of molecular oxygen (\ce{O2}) and its photochemical byproduct, ozone (\ce{O3}), are still generally regarded as important potential indicators of Earth-like life on another planet \citep{Segura2005,Kaltenegger2010,Snellen2013,Kasting2013}.

Various authors have investigated the idea that abiotic oxygen production could lead to `false positives' for life \citep{Selsis2002,Segura2007,Leger2011,Hu2013,Tian2014}. For example, it has recently been argued that the build-up of \ce{O2} to levels of $\sim2-3\times10^{-3}$ molar concentration in \ce{CO2}-rich atmospheres could occur for planets around M-class stars, because of the elevated XUV/NUV ratios in these cases \citep{Tian2014}. Extensive atmospheric \ce{O2} buildup due to \ce{H2O} photolysis followed by H escape may also occur on planets that enter a runaway greenhouse state \citep{Ingersoll1969,Kasting1988,Leconte2013b}. However, because by definition the runaway greenhouse only occurs on planets inside the inner edge of the habitable zone, it should not lead to identification of false positives for life.

For planets inside the habitable zone, it is commonly believed that \ce{H2O} photolysis will always be strongly limited by cold-trapping of water vapour in the lower atmosphere. The purpose of this note is to point out that a mechanism for \ce{O2} build-up to levels where it is the \emph{dominant} atmospheric gas exists for terrestrial\footnote{Here we define `terrestrial' in the standard (broad) way as describing any planet of low enough mass that it does not possess a dense hydrogen envelope.} planets in the habitable zone around any star type. The reason for this is that the extent of \ce{H2O} cold-trapping depends strongly on the amount of non-condensible gas in the atmosphere.

\section{Dependence of the cold trap on the non-condensible gas inventory}

Previously, we have shown that the degree to which a condensing gas such as \ce{H2O} is transported to a planet's upper atmosphere is determined primarily by the dimensionless number $\mathcal M = \epsilon p_v L \slash p_n c_p T_s$, where $L$ is the specific latent heat of the condensing gas, $c_p$ is the specific heat capacity at constant pressure of the non-condensing gas (or gas mixture), $T_s$ is temperature, $p_v$ and $p_n$ are respectively the partial pressures of the condensing and non-condensing gases in the atmosphere, $\epsilon = m_v\slash m_n$ is the molar mass ratio between the two gases, and all values are defined at the surface. $\mathcal M$ is essentially the ratio of the latent heat of the condensing gas (here, \ce{H2O}) to the sensible heat of the non-condensing gas (primarily \ce{N2} on Earth) \citep{Wordsworth2013b}. Values of $\mathcal M>1$  ($\mathcal M<1$) correspond in general to situations where the upper atmosphere is moist (dry).

\begin{figure}[h]
	\begin{center}
		{\includegraphics[width=3.0in]{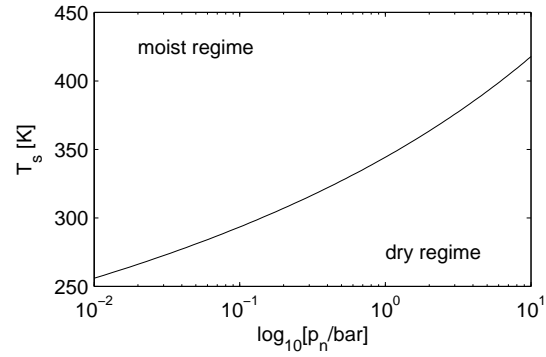}}
	\end{center}
	\caption{Surface temperature defining the transition between moist and dry upper atmosphere regimes as a function of the surface partial pressure of the non-condensable atmospheric component. Here, the non-condensing and condensing gases are \ce{N2} and \ce{H2O}, respectively. Results using \ce{O2}, \ce{Ar} or \ce{CO2} as the non-condensing gas are similar.}
	\label{fig:MeqOne}
\end{figure}

Figure~\ref{fig:MeqOne} shows the surface temperature dividing the moist and dry upper atmosphere regimes as a function of $p_n$ for a pure \ce{N2-H2O} mixture. As can be seen, on a planet with 1~bar of \ce{N2}, a surface temperature of $>340$~K is required for a moist upper atmosphere, in rough agreement with detailed radiative-convective calculations \citep{Wordsworth2013b}. However, the required surface temperature is a strong function of $p_n$. For 0.1~bar only $\sim295$~K is required, while for 0.01~bar the value drops to $\sim255$~K.
In general there is no reason to expect that Earth's atmospheric nitrogen inventory is typical for a rocky planet: in the inner Solar System alone, the range of atmospheric \ce{N2} as a function of planetary mass spans 3.3 times (Venus) to $6.6\times10^{-4}$ times (Mars) that of Earth. Delivery and removal of volatiles on terrestrial planets is dependent on an array of complex, chaotic processes, so wide variations in inventories should be expected \citep{Raymond2006,Lichtenegger2010,Lammer2009}.

\section{Abiotic oxygen on planets with pure \ce{H2O} atmospheres}

\begin{figure}[h]
	\begin{center}
		{\includegraphics[width=3.0in]{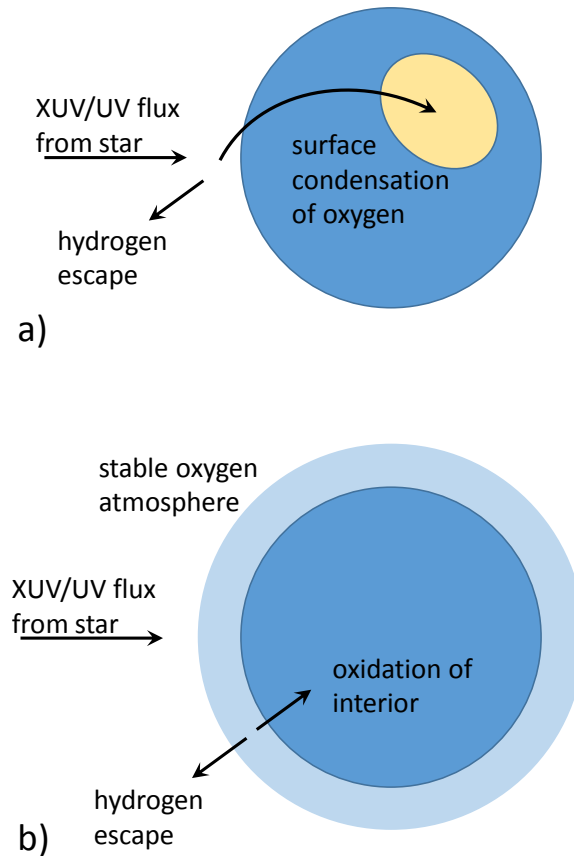}}
	\end{center}
	\caption{Schematic of possible evolutionary pathways for an initially water-dominated planet exposed to stellar XUV and UV. a) \ce{H2O} photolysis causes \ce{O2} and other oxidised products to build up on the planet's surface regions of low net instellation. b) Once sufficient \ce{O2} has built up, the planet can transition to a state where a stable \ce{O2} atmosphere is present and hydrogen escape to space is balanced by oxidation of the interior.}
	\label{fig:schematic}
\end{figure}

The \ce{O2} buildup mechanism can easily be understood intuitively by a thought experiment involving a hypothetical planet with a pure \ce{H2O} composition (Figure~\ref{fig:schematic}). Lacking atmospheric \ce{N2}, \ce{Ar} and \ce{CO2}, such a planet will initially have a pure \ce{H2O} atmosphere, with the surface pressure determined by the Clausius-Clayperon relation \citep{Andrews2010,Pierrehumbert2011b}. If the planet has the same orbit and incident stellar flux as present-day Earth, it will most likely be in a snowball state \citep{Budyko1969}. However, because \ce{H2O} cannot be cold-trapped when it is the only gas in the atmosphere, it will be photolysed by XUV and UV radiation from the host star (primarily via \ce{H2O + h\nu\to OH^* + H^*}). The resultant atomic hydrogen will escape to space at a rate dependent on factors such as the XUV energy input and the temperature of the thermosphere, and hence the atmosphere will oxidise\footnote{We assume here, as in previous work, that the efficiency of \ce{H2O} photolysis is not a limiting factor on the rate of hydrogen escape.}.

In the 1D limit with no surface mass fluxes, atmospheric \ce{O2} will build up on such a planet until $p_n$ is high enough to cold-trap \ce{H2O} and reduce loss rates to negligible values. In 3D, the initial atmospheric evolution may depend on the planet's orbit and sub-surface heat flux / transport rate, because on a tidally locked, ice-covered planet with pure \ce{H2O} atmosphere, conditions on the dark side could be so cold that even \ce{O2} would condense. However, on a planet with Earth-like rotation and obliquity, all regions of the planet receive starlight at some point in the year, so once the surface \ce{O2} inventory passed a given threshold, buildup of an \ce{O2} atmosphere would likely be inevitable. In addition, for any planet, transient heating events such as meteorite impacts would be able to force transitions to a stable state of high atmospheric pressure\footnote{The latent heat of sublimation of \ce{O2} ($L_\ce{O2}=213$~kJ~kg$^{-1}$) is only around 1/10\textsuperscript{th} that of \ce{H2O} \citep{CRC2000}. Hence with only 25\% energy conversion efficiency, the kinetic energy of an impactor travelling at 10~km~s$^{-1}$ with density 3~g~cm$^{-3}$ would be sufficient to sublimate a 1-bar atmosphere of \ce{O2} on an Earth-size planet if its radius was 19.2~km.}.

What about more general scenarios? First, we can relax the assumption of zero downward flux at the surface and consider cases where the created \ce{O2} can be used to oxidise the interior. Then, redox balance dictates that atmospheric oxygen levels must build up until the loss of hydrogen to space is balanced by the surface removal rate of oxidising material. For example, if an \ce{O2} removal
rate\footnote{The actual rate of interior oxidation of an \ce{H2O} world with an oxygen-rich atmosphere is difficult to calculate. For comparison, the average rate of oxidation due to \ce{Fe^{3+}} subduction to the mantle on Earth over the last 4~Gy was estimated as $1.9-7.1\times10^{9}$~molecules~\ce{O2}~cm$^{-2}$~s$^{-1}$ in \cite{Catling2001}.}
of $5\times10^9$~molecules~cm$^{-2}$~s$^{-1}$ at the surface is balanced by diffusion-limited \ce{H2O} loss, given an escape rate $\Phi = b_\ce{H2O-O2}f_\ce{H2O} (H_\ce{O2}^{-1}-H_\ce{H2O}^{-1})$, the molar concentration of \ce{H2O} at the cold trap\footnote{The relationship between $\Phi$ and $f_\ce{H2O}$ depends weakly on the homopause temperature $T_h$ via the scale heights and $b_\ce{H2O-O2}$. For simplicity, $T_h=300$~K is used here.} must be $3\times10^{-3}$~mol/mol under Earth gravity. Here $b_\ce{H2O-O2}$ is the binary diffusion coefficient of \ce{H2O} in \ce{O2} \citep{Marrero1972}, $H_\ce{O2}$ and $H_\ce{H2O}$ are respectively the atmospheric scale heights of \ce{O2} and \ce{H2O}, and $f_\ce{H2O}$ is the cold trap \ce{H2O} molar concentration.

The surface \ce{O2} partial pressure required to match this cold-trap concentration, which can be calculated by integrating the moist adiabat equation \citep{Ingersoll1969} as in \cite{Wordsworth2013b}, depends on both the surface and cold-trap temperatures. In Earth's present-day oxygen-rich atmosphere, the cold trap occurs at a relatively high $T_t\sim210$~K, due primarily to the warming effect of ultraviolet solar absorption by \ce{O3} \citep{Andrews2010}. Given $T_s = 288$~K and $T_t = 210$~K, $f_\ce{H2O}  = 3\times10^{-3}$~mol/mol requires a surface \ce{O2} partial pressure of 0.15~bar. For a snowball planet with $T_s = 240$~K, this would drop to 0.022~bar. By comparison, for $T_s=288$~K and $T_t=140$~K, 0.025~bars is required\footnote{The $T_t=140$~K calculation may underestimate the required surface \ce{O2} partial pressure, because effective blocking of \ce{H2O} photolysis also requires the cold-trap altitude to be lower than that at which the atmospheric opacity in the UV becomes less than unity.}. Because \ce{O2} build-up should lead to \ce{O3} formation and hence stratospheric heating, \ce{O2} partial pressures of at least a fraction of a bar appear plausible once the planet's atmosphere reaches a steady state.

\section{Abiotic oxygen on Earth-like planets}

How would things change on a more complex planet where other atmospheric constituents were present? First, if the atmosphere contains some \ce{N2} or \ce{Ar}, the amount of \ce{O2} required to block \ce{H2O} escape will clearly be decreased, and increased horizontal heat transport would reduce the likelihood of atmospheric bistability via \ce{O2} condensation in the planet's regions of low surface instellation. Reduced gases such as methane, which can be outgassed from a planet's interior by abiotic processes \citep{Levi2013,Guzman2013}, could have lifetimes similar to those on Earth today in an \ce{O2}-rich atmosphere, although variations in \ce{O3} and \ce{NO_x} concentrations as a function of UV levels and atmospheric composition might alter this \citep{Wayne2000}. In addition, volcanically emitted sulphur species and heterogenous chemistry will also affect the atmospheric redox balance. Future investigations using photochemistry models will allow constraints on the importance of these effects as a function of the water loss rate.

Surface/interior redox exchanges are another source of complexity on a low-\ce{N2} Earth-like planet.
If the planet forms with a hydrogen envelope that is lost to space early on \cite[e.g., ][]{Genda2008}, its crust and oceans should initially be reducing, and the oxidised products of \ce{H2O} photolysis might react rapidly with the surface at first. However, as long as this occurred, the upper atmosphere would remain \ce{H2O}-rich and rapid photolysis could continue. Over time, the planetary surface and interior would become oxidised, decreasing their ability to act as an oxygen sink. Assuming Earth's present-day XUV flux, a lower limit on \ce{H2} escape from a hydrogen-rich homopause is $\sim 4\times10^{10}$~molecules~cm$^{-2}$~s$^{-1}$ \citep{Tian2005}. Given this, an \ce{N2}-poor Earth could lose $2.1\times10^{22}$~moles of \ce{H2O} over 4~Gy, or 28\% of the current ocean volume\footnote{In this calculation, we assume that 50\% of the escaping hydrogen is outgassed directly from the mantle.}. This translates to 66.2~bar of atmospheric \ce{O2} -- a large enough quantity to cause significant irreversible oxidation of the solid planet and hence a strong decrease in the reducing power of the surface. Because XUV fluxes are greatly enhanced around young dwarf stars in general, total water loss could be many times this value in many cases \citep{Ribas2005,Ribas2010,Linsky2014}.

Finally, an Earth-like planet could have \ce{CO2} outgassing, plate tectonics and hence the potential for a carbonate-silicate weathering feedback \citep{Walker1981}. The \ce{CO2} cycle on an initially anoxic planet without \ce{N2} or \ce{Ar} would be complex, because \ce{CO2} condenses at relatively high temperatures \citep{CRC2000} but has low compressive strength in solid form \citep{Clark1976}. In the absence of ocean/interior heat transport processes, outgassed \ce{CO2} could build up on the low instellation regions of a planet until the return flow of \ce{CO2} glaciers became sufficient to transport it back to high instellation regions.

\begin{figure}[h]
	\begin{center}
		{\includegraphics[width=3.0in]{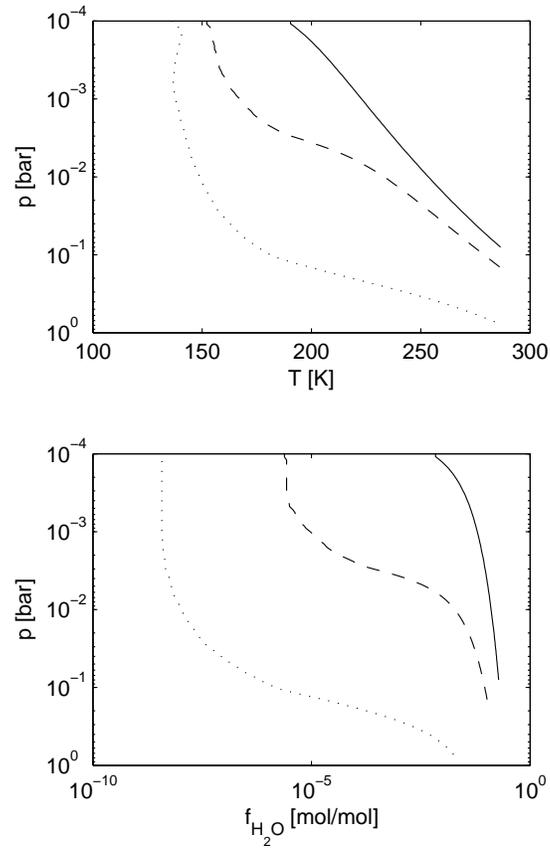}}
	\end{center}
	\caption{Atmospheric a) temperature and b) \ce{H2O} molar concentration in thermal equilibrium as a function of pressure, as simulated by the 1D radiative-convective model. In each case the atmospheric composition is \ce{N2-CO2-H2O}. For the dotted, dashed and solid lines, the \ce{N2} inventories are 1, 0.17 and 0.007 times that of present-day Earth, and the dry \ce{CO2} molar concentration is $1\times10^{-3}$, 0.1 and 0.9~mol/mol, respectively. As can be seen, the upper atmosphere is moist when \ce{N2} levels are low, implying rapid \ce{H2O} photolysis.}
	\label{fig:profiles}
\end{figure}

Setting aside the complexity of the full climate problem for future study, we can nonetheless demonstrate the potential for \ce{O2} build-up in cases where \ce{CO2} levels are such that the planet has an Earth-like global mean surface temperature. Figure~\ref{fig:profiles} shows the variation of atmospheric temperature and \ce{H2O} molar concentration with atmospheric \ce{N2} content calculated using the same methodology as in \cite{Wordsworth2013b}, for an Earth-like planet at 1~AU around a Sun-like star, assuming an \ce{N2-CO2-H2O} atmosphere with tropospheric \ce{H2O} relative humidity of 0.5. In each case, the \ce{CO2} molar concentration has been chosen to yield close to $T_s = 288$~K in equilibrium.  As can be seen, once the \ce{N2} content drops below a few percent of that on present-day Earth, the high atmosphere becomes rich in \ce{H2O}, implying rapid photolysis and hence planetary oxidation. Hence we may conclude that even planets that are Earth-like in all respects except for the \ce{N2} content of their atmospheres have the potential to build up \ce{O2} abiotically until it is a major atmospheric constituent.

\section{Conclusion}

Because \ce{O2} can become the dominant gas in the atmosphere of a lifeless planet, alone it cannot be regarded as a robust biosignature. Our results do not necessarily rule out its utility in every case. However, they do demonstrate that the situation is considerably more complex than has previously been believed, with the likelihood of an abiotic \ce{O2}-rich atmosphere emerging a complicated function of a planet's accretion history, internal chemistry, atmospheric dynamics and orbital state. Investigation of the range of possibilities for terrestrial planets with variable \ce{N2} and noble gas inventories should be a rich area for future theoretical research that will help to expand our understanding of climate evolution mechanisms. Nonetheless, for a specific exoplanet, even detailed modelling might not lead to a definite conclusion given the inherent uncertainties in processes such as volatile delivery during formation.

Observationally, there may still be a way to distinguish the scenarios we discuss here, but only if a reliable way is developed to retrieve the ratio of \ce{O2} to \ce{N2} or \ce{Ar} in an exoplanet's atmosphere.  In principle this may be achieved by analysis of the planet's spectrally resolved phase curve \citep{Selsis2011}, or in transit by measurement of the spectral Rayleigh scattering slope \citep{Benneke2012} in a clear-sky (i.e., aerosol-free) atmosphere, or possibly via spectroscopic observation of oxygen dimer features \citep{Misra2014}. More work will be required to assess the potential of these techniques to determine \ce{O2}/\ce{N2} mixing ratios in realistic planetary atmospheres.

\acknowledgments

R.~W. acknowledges support from the National Science Foundation and NASA's VPL program. This article benefited from discussions with F.~Tian, R.~de~Kok, S.~Rugheimer and D.~Sasselov.


\end{document}